\documentclass[twocolumn,superscriptaddress,showpacs,pra]{revtex4}
\usepackage{epsfig}
\usepackage{epstopdf}
\usepackage{delarray}
\usepackage{amsmath, amssymb}
\usepackage{bm}
\usepackage{graphicx}
\usepackage{placeins}
\usepackage{color}

\begin{document}
\title{Rogue Wave Spectra of the Kundu-Eckhaus Equation}

\author{Cihan Bay\i nd\i r}
\email{cihan.bayindir@isikun.edu.tr}
\affiliation{Engineering Faculty, I\c{s}\i k University, \.{I}stanbul, Turkey}

\begin{abstract}
In this paper we analyze the rogue wave spectra of the Kundu-Eckhaus equation (KEE). We compare our findings with their nonlinear Schr\"{o}dinger equation (NLSE) analogs and show that the spectra of the individual rogue waves significantly differ from their NLSE analogs.  A remarkable difference is the one-sided development of the triangular spectrum before the rogue wave becomes evident in time. Also we show that increasing the skewness of the rogue wave results in increased asymmetry in the triangular Fourier spectra. Additionally, the triangular spectra of the rogue waves of the KEE begin to develop at earlier stages of the their development compared to their NLSE analogs, especially for larger skew angles. This feature may be used to enhance the early warning times of the rogue waves.  However we show that in a chaotic wavefield with many spectral components the triangular spectra remains as the main attribute as a universal feature of the typical wavefields produced through modulation instability and characteristic features of the KEE's analtical rogue wave spectra may be suppressed in a realistic chaotic wavefield.

\end{abstract}
\maketitle


\section{\label{sec:level1} Introduction}
Rogue (freak) waves may be detected by spectral analysis before they become evident in time \cite{Akhmediev2011, bayindir2016}. Although some recent attempts try to measure the wavefield directly in the spatial domain using efficient signal processing techniques \cite{bayindir2016arxivcsearly}, spectral analysis is still the main tool especially for the ultrafast optic studies \cite{Hollas, Akhmediev2015PhysD}.  

Exact rogue wave solutions of the different integrable systems differ significantly in shape \cite{Akhmediev2015PhysD}. Therefore it is natural to expect that their spectral features, possible early detection mechanisms and times may differ as well. In this work we show that the spectra of the individual rogue wave solutions of the Kundu-Eckhaus (KEE) differ significantly from those of the nonlinear Schr\"{o}dinger equation (NLSE). Although the spectra of the individual rogue waves are significantly different, the rogue wave spectra of the chaotic wavefields have certain similarities with the NLSE case.
 
The Kundu-Eckhaus equation (KEE) in one of the integrable extensions of the NLSE \cite{bayindir2016PRE, bayindir2016KEEpot}. It contains extension terms to the standard cubic NLSE, namely the quintic and Raman-effect nonlinear terms \cite{Wang, Zhao2013, dqiu}. One of the different versions of the KEE can be written in the form of
\begin{equation}
i\psi_t + \psi_{xx} + 2 \left|\psi \right|^2 \psi + \beta^2 \left|\psi \right|^4 \psi - 2 \beta i \left( \left|\psi \right|^2 \right)_x \psi =0
\label{eq01}
\end{equation}
where $\psi$ is complex amplitude, $x,t$ are the spatial and temporal variables and $i$ is the imaginary number \cite{Wang}. The $\beta$ parameter is a real constant and $\beta^2$ is the coefficient of the quintic nonlinear term. The last term of the KEE represents the Raman-effect which accounts for the self-frequency shift of the waves \cite{Wang}. KEE equation can adequately model the propagation of the ultrashort pulses in nonlinear and quantum optics, which can possibly be used to describe the optical properties of the femtosecond lasers and can be used in femtochemistry studies. Some extensions of the NLSE, similar to the form of the KEE, where quintic nonlinearity is not included but third order dispersion and gain and loss terms are included are also used as models in the soliton-similariton laser studies \cite{Ilday}.

Some analytical periodic and rational solutions of the KEE given in Eq.(\ref{fig1}) exist in the literature \cite{Wang, dqiu}. The first order rational solution of the KEE is given by
\begin{equation}
\psi_1=\exp{\left[i (- \beta x+ (\beta^2+2)t) \right]} \frac{L_1+i J_1}{M_1} \exp{ \left[i \beta \frac{K_1}{M_1}  \right]}
\label{eq02}
\end{equation}
where 
\begin{equation}
\begin{split}
& L_1=-4x^2-16 \beta t x-16(\beta^2+1)t^2+3, \ \ \ J_1=16t, \\
& M_1=4 x^2+ 16 \beta t x + 16(\beta^2+1)t^2+1, \\
& K_1=4 x^3+ 16 (\beta^2+1) t^2 x + 9x+ 16\beta(x^2+1)t.
\label{eq03}
\end{split}
\end{equation}
This solution and some other analytical solutions are given in \cite{Wang, dqiu, Zhao2013}. This first order rational rogue wave is basically a skewed Peregrine soliton of the NLSE. Setting the parameter $\beta=0$, the KEE reduces to the cubic NLSE for which the rogue wave solutions become obvious and become the rational soliton solutions of the NLSE \cite{Akhmediev2009b}. For the cubic NLSE, the first and higher order rational rogue wave solutions can be seen in \cite{Akhmediev2009b}.  Second and the higher order rational solutions of the KEE and a hierarchy of obtaining those rational solutions based on Darboux transformations are presented in \cite{Wang}. They are basically skewed rogue waves obtained by the gauge transforming the rogue wave solutions of the cubic NLSE. For the sake of brevity, we are not repeating their explicit formulas here. For the details of their formulation the reader is referred to \cite{Wang}.

\section{\label{sec:level1} Spectra of Individual Rogue Waves} 
In order to analyze the properties of the rogue wave spectra in a chaotic wave field, first we should analyze the spectra of the individual rogue wave solutions of the KEE. We obtain the spectra by Fourier transform operation. Throughout this paper we denote by the letter F the spectra obtained by the Fourier transform of the wavefield $\psi(x,t)$ in the spatial variable $x$, i.e.
\begin{equation}
F(k,t)=\int_{-\infty}^{\infty} \psi(x,t) \exp(-ikx)dx 
\label{eq04}
\end{equation}
where $k$ denotes the wavenumber parameter. The spectra calculated this way are complex functions and we present the energy spectral density, $ \left|F(k,t)\right|^2$ in all of the spectra calculations throughout this paper. Energy spectral density is a parameter which can be measured directly in experimental optics \cite{Akhmediev2015PhysD}.
\begin{figure}[h]
\begin{center}
   \includegraphics[width=3.5in,height=4.1in]{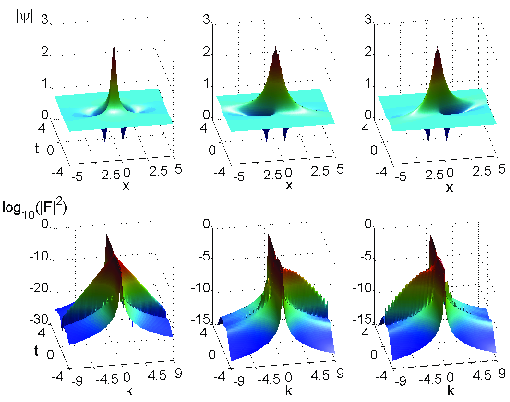}
  \end{center}
\caption{\small First row from left to right: First order rogue wave of the KEE for $\beta=0, 1, -1$, respectively. Second row: Corresponding spectra in logarithmic scale.}
  \label{fig1}
\end{figure}

\begin{figure}[h]
\begin{center}
   \includegraphics[width=3.5in,height=4.1in]{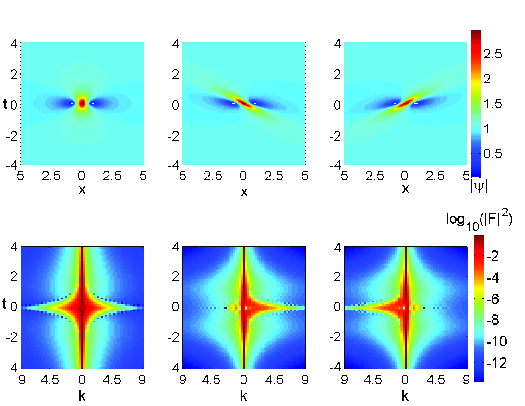}
  \end{center}
\caption{\small First row from left to right: Color contour plot of the first order rogue wave of the KEE for $\beta=0,1,-1$, respectively. Second row: Color contour plot of the corresponding spectra in logarithmic scale.}
  \label{fig2}
\end{figure}

The first order rational rogue wave solutions of the KEE and the corresponding spectra are shown in Fig.\ref{fig1} for $\beta=0,1,-1$, respectively. First 3D plot in the second row of Fig.\ref{fig1} shows the Fourier spectrum of the Peregrine soliton. The analytical form of this spectrum is given in \cite{Akhmediev2011exactpspectra} and for the sake of brevity it will not be repeated here. We compute the Fourier transform of the first order rogue wave solution of the KEE given by Eq. \ref{eq02} numerically by using fast Fourier transform routines. 3D plots of spectra of the first order rational soliton solution of the KEE are also presented in the second and third places in second row of the Fig. \ref{fig2} for $\beta=1,-1$, respectively. The results depicted in Fig. \ref{fig1} are also given as color contour plots in Fig. \ref{fig2} for a better visualization of the spectra properties. 

\begin{figure}[h!]
\begin{center}
   \includegraphics[width=3.5in,height=4.1in]{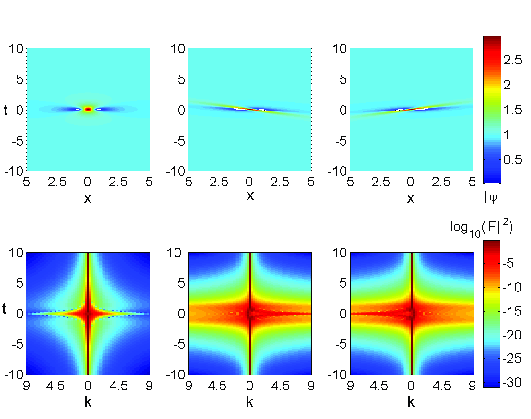}
  \end{center}
\caption{\small First row from left to right: Color contour plot of the first order rogue wave of the KEE for $\beta=0, 1.75, -1.75$, respectively. Second row: Color contour plot of the corresponding spectra in logarithmic scale.} \label{fig3}
\end{figure}

As discussed in \cite{bayindir2016PRE, Wang} we can see that quintic and Raman-effect nonlinear terms produce an important skew angle relative to the ridge of the rogue waves. The sign of the $\beta$ parameter determines the skewness direction relative to the ridge of the rogue wave \cite{bayindir2016PRE, Wang}. If $\beta = 0$, then there is no skewness and the rogue wave solution of the KEE reduces to the Peregrine soliton solution of the NLSE. For $\beta > 0$, the skewness is in the counter clockwise direction whereas for $\beta < 0$ it is in the clockwise direction \cite{bayindir2016PRE, Wang}. Checking Fig. \ref{fig2} and Fig. \ref{fig3}, the distinct feature of the rogue wave spectra of the KEE compared to their NLSE analogs is that, the spectra are strongly asymmetric. For a skewness in the counter clockwise direction, which occurs due to the positive $\beta$ parameter, the triangular widening occurs in the positive wavenumbers. For negative values of the $\beta$ parameter, this situation reverses and triangular widening becomes apparent in the negative wavenumbers. For both of the cases triangular widening is clearly distinguishable in only one side of the spectra.

Next we analyze the effect of increasing the skewness on the spectral features of the rogue waves of the KEE. For this purpose we set $\beta=0, 1.75, -1.75$ and depict the corresponding spectra in Fig. \ref{fig3}. As can be seen from the contour plots given in Fig. \ref{fig3}, as the skewness of the rogue wave increases due to larger  values of $\left|\beta\right|$, the triangular widening of the spectra becomes more significant. Additionally, the asymmetry in the spectra becomes more prominent as well. For larger skewness, the higher absolute wavenumber components begin to acquire some energy and they deviate from zero at earlier times of the rogue wave emergence compared to their NLSE analogs. For example for $t \approx -5$, the first order rational solution of the KEE has significantly more widening due to energy acquired by the higher absolute wavenumbers compared to its NLSE analog. Even for $t \approx -7$, the widening of the spectra of the KEE is larger than those of the NLSE. This feature of the KEE spectra may be used to enhance the early detection times of the rogue waves.

\begin{figure}[h!]
\begin{center}
   \includegraphics[width=3.5in,height=4.1in]{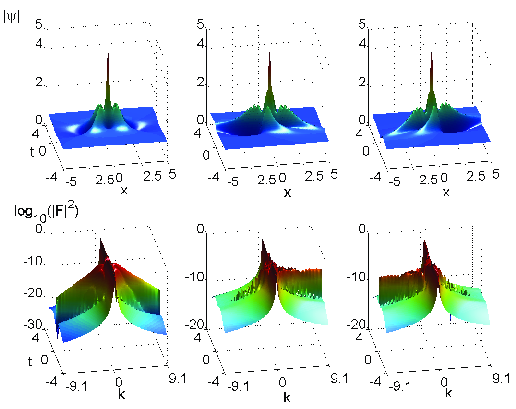}
  \end{center}
\caption{\small First row from left to right: Second order rogue wave of the KEE for $\beta=0, 1, -1$, respectively. Second row: Corresponding spectra in logarithmic scale.}
  \label{fig4}
\end{figure}

\begin{figure}[h!]
\begin{center}
   \includegraphics[width=3.5in,height=4.1in]{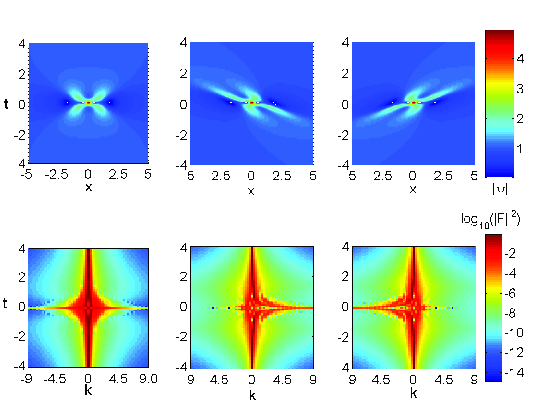}
  \end{center}
\caption{\small First row from left to right: Color contour plot of the second order rogue wave of the KEE for $\beta=0, 1, -1$, respectively. Second row: Color contour plot of the corresponding spectra in logarithmic scale.} \label{fig5}
\end{figure}

\begin{figure}[h]
\begin{center}
   \includegraphics[width=3.5in,height=4.1in]{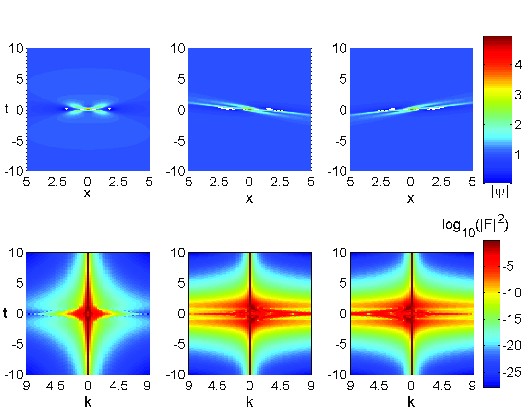}
  \end{center}
\caption{\small First row from left to right: Color contour plot of the second order rogue wave of the KEE for $\beta=0, 1.75, -1.75$, respectively. Second row: Color contour plot of the corresponding spectra in logarithmic scale.} \label{fig6}
\end{figure}

We also analyze the second order rogue wave spectra of the KEE. For the sake of the brevity we do not repeat the exact formulation of the second order rational soliton solution of the KEE. It is given by Eq. 30 of \cite{Wang}. We confine ourselves with presentation of their numerical transforms. Corresponding results are depicted in Figs.\ref{fig4}-\ref{fig6}. The results are quite similar to the results obtained for the first order rogue wave of the KEE, however there are more dips in the spectra due to increased number of zeros in the absolute value of the wavefunction ($\left|\psi\right|$). Similar to the first order case, the skewness in the second order rogue wavefield causes a strong asymmetry in the spectra. The triangular widening occurs distinctly in positive wavenumber side of the spectrum for the positive values of the $\beta$ parameter and in the negative wavenumber side of the spectrum for the negative values of the $\beta$ parameter. Similar to the first order rogue wave case, increasing the skewness in the wave profile results in more asymmetry in the rogue wave spectra which begin to develop at earlier times of rogue wave emergence compared to their NLSE analogs. 

\noindent For example for $t \approx -5$, there is a significant difference in the widening of the triangular spectra of the second order rogue wave of the KEE compared to its NLSE analog shown in Fig.\ref{fig6}. For $t \approx -7$, the difference begins to become significant. Using this feature, the individual rogue waves of the KEE can be detected at earlier stages of their development compared to their NLSE analogs by spectral measurements in practice. This may be done for a wavefield under quintic and Raman-effect nonlinear effects. While in an optical setup this may be performed by adjusting the optical properties of the medium, in hydrodynamics a realization would be extremely difficult and only naturally emerging skewed rogue waves may give some clue about the usability of this feature. However we can not answer the questions about details of the applicability of this feature in this study. Since a realistic wavefield would include many spectral components, we turn our attention to analyze the chaotic wavefields with many spectral components generated in the frame of the KEE by the modulation instability.

\section{\label{sec:level1}Spectra of the chaotic wave field}

The processes modeled in the frame of the partial differential equations such as KEE can be very complicated. However they are still governed by a deterministic equation. Therefore their results can be predicted for a given initial condition. Therefore compared to the completely unpredictable  'stochastic' processes, the processes described in the frame of the KEE  can be described as 'chaotic' \cite{Akhmediev2014}.  The term 'chaotic' is used in this setting throughout this paper. We use a numerical framework in order to analyze chaotic wavefields in the frame of the KEE. We start the wavefield simulations using a constant amplitude sinusoid with an additive small amplitude white noise. Such a state is unstable and it evolves into a full-scale chaotic wave field similar to the numerical results given in \cite{Akhmediev2009b, Akhmediev2009a, bayindir2016}. The chaotic wave field modeled by the KEE with this starter evolves into a wavefield which exhibits many amplitude peaks, with some of them becoming rogue waves. This behavior is similar to the results obtained for the NLSE and Sasa-Satsuma equations (SSE) \cite{Akhmediev2015PhysD}. In order to model such a chaotic wavefield starter, we use the initial condition
\begin{equation}
\psi (x,t=0)=\psi_0 (x,0)+\mu a(x)
\label{eq08}
\end{equation}
where $\psi_0=\exp(i k_0 x)$ initial plane wave solution,  $k_0$ is the initial seed plane wavenumber which is selected as $k_0=0.1$. $a(x)$ is a uniformly distributed random complex function with real and imaginary parts having random values in the interval of $[-1,1]$. Following \cite{Akhmediev2009b, Akhmediev2009a, bayindir2016}, a value of $\mu=0.2$ is selected. The water surface fluctuation would be given by the real part of $ \left|\psi\right| \exp{[i\omega t]}$ where $\omega$ is a carrier wave frequency however the parameter we investigate is the envelope ($ \left|\psi\right|$) of the chaotic wavefield. For the time integration of the KEE, we use the split-step Fourier scheme described in \cite{bayindir2016PRE}. Briefly, in typical split-step Fourier schemes the spatial derivatives are evaluated using spectral techniques that employ Fourier transforms in periodic domains \cite{bayindir2016scirep, bayindir2009, Karjadi2010, Karjadi2012, trefethen, demiray, bayindir2015arxivcsmww, bayindir2015d, bayindir2015arxivchbloc, bayindir2015arxivcssfm, bayin2015b, bayin2015c, bayin2015e} and time stepping is performed by an exponential function. For the sake of brevity, we will not include the details of the split-step scheme of the KEE here. The reader is referred to the \cite{bayindir2016PRE} for a more detailed explanation.

\begin{figure}[h]
\begin{center}
   \includegraphics[width=3.5in,height=4.1in]{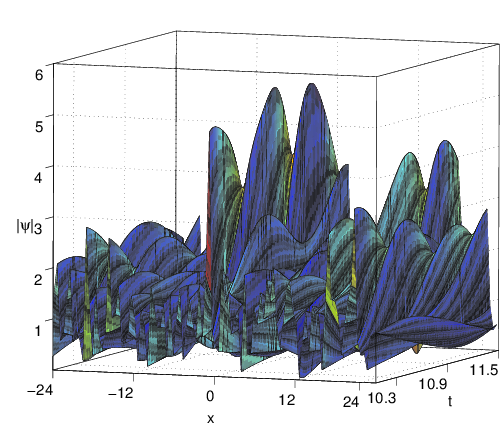}
  \end{center}
\caption{\small An example of a rogue wave in the chaotic wavefield. The amplitude of this wave exceeds 5.} \label{fig7}
\end{figure}

Modulation instability is one of the methods used to generate a chaotic wavefield \cite{Akhmediev2015PhysD}. Modulation instability started by the noise formulated above creates a chaotic wavefield that starts from the initial plane-wave. As recently discussed in \cite{bayindir2016PRE}, the chaotic wavefield of the KEE is skewed in the counter clockwise direction for $\beta > 0$ and it is skewed in the clockwise direction for $\beta < 0$, similar to the analytical results. Therefore sign of the $\beta$ parameter controls the skewness direction of the wavefield but it does not affect the probability of rogue wave occurrence \cite{bayindir2016PRE}. Filaments of the chaotic wavefield propagate approximately with the average group velocity \cite{bayindir2016PRE}. Additionally, as the values of the initial seed plane wavenumber ($k_0$) gets smaller, the probability of occurrence of extreme waves in the chaotic wavefield increases \cite{bayindir2016PRE}. Similar behavior is also observed for the NLSE and SSE \cite{Akhmediev2015PhysD}, which is possibly an indicator of universal property of the processes started with modulation instability.

\begin{figure}[h]
\begin{center}
   \includegraphics[width=3.5in,height=3.6in]{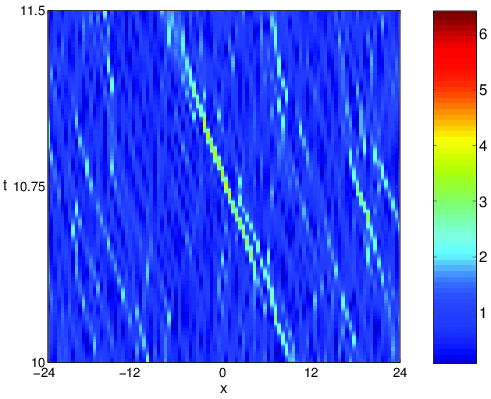}
  \end{center}
\caption{\small Contour plot of the rogue wave shown in Fig.\ref{fig7}.} \label{fig8}
\end{figure}

In order to demonstrate the possible usage and examine the features of the spectra for the early detection of the rogue waves of the KEE, we choose an area of the chaotic wavefield with significantly higher amplitude than other parts. An example of a patch of the chaotic wavefield with a rogue wave exceeding amplitude $5$ at $t \approx 10.8$ is shown in Fig. \ref{fig7} in 3D format. The same chaotic wavefield is shown in contour map format in Fig. \ref{fig8}. A value of $\beta = 1.75$ is used in this simulation and the wavefield shown in Fig. \ref{fig8} is skewed to the left due to positive value of the $\beta$ parameter, as discussed in \cite{bayindir2016PRE}.

The spectrum of the chaotic wavefield containing the rogue wave is shown in Fig. 9. In order to isolate the features due to the rogue wave in the presence of many spectral components, the spectrum is obtained after a super (flat-head)-Gaussian mask centered at the location of the peak amplitude is applied to the chaotic wavefield. That is, the mask is applied by simple multiplication of the super-Gaussian function with the chaotic wavefield which includes the above mentioned rogue wave in the physical domain. If whole of the rogue wave is within the masked zone, than centering of this super-Gaussian function does not affect the spectral results significantly \cite{Akhmediev2015PhysD}.

\begin{figure}[h]
\begin{center}
   \includegraphics[width=3.5in,height=4.1in]{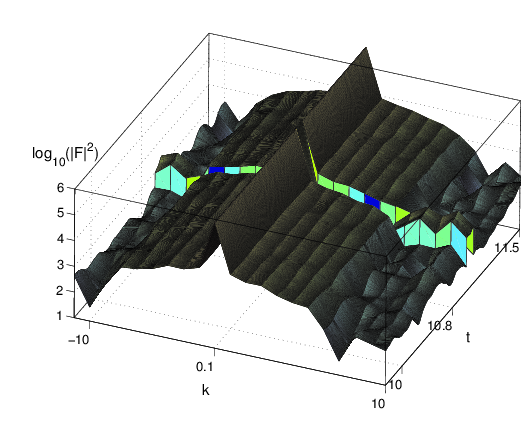}
  \end{center}
\caption{\small The spectrum of the chaotic wavefield shown in Fig.\ref{fig7}. The apparent widening of the spectrum at $t\approx 10.8$ is due to appearance of a rogue wave.} \label{fig9}
\end{figure}

The spectrum of the patch of the wavefield shown in Fig. 9 has a visible widening due to appearance of the rogue wave. A point-by-point comparison of this spectrum with the spectra of the analytical solutions of the KEE presented in the previous sections is extremely difficult. Although some asymmetry in the triangular spectrum with more energy in the positive wavenumbers due to the counter clockwise skewness of the wavefield which is due to the positive $\beta$ parameter can be observed, this is not the prominent feature of the spectrum.  The prominent feature of the spectrum displayed in Fig. 9 is the considerable widening of the spectrum due to the emergence of the rogue wave. This feature can be used to reveal rogue wave emergence from spectral measurements \cite{Akhmediev2015PhysD}. In reality it is only possible to measure the part of the wavefield. However similar to the NLSE and the SSE cases, the spectrum remains triangular even if it is calculated for the whole or a part of the chaotic wavefield \cite{Akhmediev2015PhysD}. This is the main attribute as a universal feature of the typical chaotic wavefields produced through modulation instability \cite{Akhmediev2015PhysD, Akhmediev2011untrian} and characteristic features of the KEE's analytical rogue wave spectra may be suppressed in a realistic chaotic wavefield.

\section{\label{sec:level1}Conclusion}

In this paper, we have studied the spectral features of the first and second order rational rogue wave solutions of the Kundu-Eckhaus equation. Individual spectra of the rogue waves of the Kundu-Eckhaus equation significantly differ from their NLSE analogs. They exhibit strong asymmetry due to one-sided development of the triangular spectra before the rogue waves become evident in time. As the skewness of the wave field, which is controlled by the $\beta$ parameter, increases; so does the asymmetry in the triangular spectra. Additionally the development of the triangular spectra of the rogue waves of the Kundu-Eckhaus equation occur before their NLSE analogs, which may be used to enhance the early warning times.  However the rogue wave spectra of a choatic wave field studied in the frame of Kundu-Eckhaus equation have similar triangular widening signature with the NLSE case as a universal feature of the fields resulted from modulation instability. Therefore characteristic features of the analytical rogue wave spectra of the Kundu-Eckhaus equation may be suppressed in a realistic chaotic wavefield.

\end{document}